\documentstyle[aps,preprint]{revtex}
 \tightenlines
\begin{document}
\draft
\title{A Cellular Automaton Model for Diffusive and Dissipative Systems}
\author{T. C. Chan$^{1}$, H. F. Chau$^{2,3}$\footnote{To whom all
 correspondence should be addressed.}, and K. S. Cheng$^{1}$}
\address{
{}~$^1$ Department of Physics, University of Hong Kong, Pokfulam Road,
 Hong Kong.\\
{}~$^2$ School of Natural Sciences, Institute for Advanced Study, Olden Lane,
 Princeton, NJ 08540, USA.\\
{}~$^3$ Department of Physics, University of Illinois, 1110 West Green Street,
 Urbana, IL 61801, USA.
}
\date{\today}
\preprint{IASSNS-HEP-94/80; cond-mat:9410007}
\maketitle
\mediumtext
\begin{abstract}
We study a cellular automaton model, which allows diffusion of energy (or
equivalently any other physical quantities such as mass of a particular
compound) at every lattice site after each timestep.  Unit amount of energy
is randomly added onto a site. Whenever the local energy content of a site
reaches a fixed threshold $E_{c1}$, energy will be dissipated. Dissipation of
energy propagates to the neighboring sites provided that the energy contents
of those sites are greater than or equal to another fixed threshold $E_{c2}
(\leq E_{c1})$. Under such dynamics, the system evolves into three different
types of states depending on the values of $E_{c1}$ and $E_{c2}$ as reflected
in their dissipation size distributions, namely: localized peaks, power laws,
or exponential laws. This model is able to describe the behaviors of various
physical systems including the statistics of burst sizes and burst rates in
type-I X-ray bursters. Comparisons between our model and the famous
forest-fire model (FFM) are made.
\end{abstract}
\medskip
\section{Introduction}
For many years, cellular automaton models have been playing important roles
in the study of some non-linear and spatially extended systems. In most cases,
these models cannot be solved analytically and their investigations rely
mainly on computer simulations. Forest-fire model (FFM) \cite{Bak90}, Eden
model \cite{Eden} and earthquake simulations \cite{EQ} are some examples.
Besides, a lot of simplifications has been taken in these models, making their
underlying physics unclear. In this paper, we introduce a simple cellular
automaton model that accounts for the stochastic energy (or equivalently any
other physical quantity) introduction, occasional energy dissipation, and
energy diffusion in an open physical system. This kind of non-equilibrium
systems are common in nature. We find that the system shows different
behaviors with different parameters. And with an appropriate choice of
parameters, the model can be applied (sometimes after a small changes in the
geometry) to explain the statistics of the occasional outburst of energy (or
other physical quantities) in various physical systems including the
thermonuclear runaways in X-ray bursters \cite{Chau94} and the sudden CO$_2$
gas release in crater lakes \cite{Lakes}.
\par \medskip
We introduce the model in Section~II. Then we discuss the significance of
various parameters and present the general results of computer simulation in
Section~III. In Section~IV, we concentrate on a special case and compare it
with the FFM. Finally, a conclusion together with a discussion on the
application of our model can be found in Section~IV.
\section{The Model Of Diffusion And Dissipation}
We consider a $L \times L$ square lattice each of size $\Delta L\times \Delta
L$. The energy contained in a lattice site $\bf r$ is denoted by $E({\bf r})$,
which is a non-negative real number. Since the sites are of equal areas, the
ratio of energy to energy density is a constant for all site. Evolution of
the system is governed by the following rules:
\begin{enumerate}
\item {\em Energy Introduction:} at each timestep, a unit amount of energy is
added to a randomly selected site $\bf r_0$, i.e.
\begin{equation}
 E({\bf r_0})\rightarrow E({\bf r_0})+1 \mbox{.}
\end{equation}
It should be noted that the energy introduction rate to the entire system is
being fixed. Our energy introduction rule takes only spatial fluctuation into
account.
\item {\em Triggering Of Energy Dissipation:} whenever the energy of a site
$\bf r_0$ reaches a fixed value $E_{c1}$ called the triggering threshold,
energy stored at that site will be dissipated, i.e.
\begin{equation}
 \mbox{if }E({\bf r_0}) \ge  E_{c1}\mbox{ ,then } E({\bf r_0})\rightarrow 0
 \mbox{.}
\end{equation}
Note that the triggering threshold $E_{c1}$ is the same at all sites and is
time independent. The dissipation process is called ``burning''. Since all
sites are of equal areas, we can regard the ``fire'' as being triggered
whenever the local energy density at some place exceeds a predetermined
triggering threshold.
\item {\em Propagation Of Energy Dissipation:} nearest neighbors of a burning
site burn provided that their energy content are greater than or equal to a
fixed value $E_{c2}$ ($\le E_{c1}$) called the propagation threshold, i.e.
\begin{equation}
 \mbox{if }E({\bf r_0'}) \ge E_{c2}\mbox{ ,then } E({\bf r_0'})\rightarrow 0
\end{equation}
whenever $\bf r_0'$ is a nearest neighbor of the burning site $\bf r_0$. We
use periodic boundary conditions in the determination of neighboring sites.
One may regard the triggering threshold $E_{c1}$ as the (energy) density
required to activate the dissipative process. Once the ``burning'' takes
place, the energy release may heat up the neighboring sites and hence the
density threshold required for the fire to propagate, $E_{c2}$, is reduced.
This is valid in many physical and chemical reactions.
\par
The burning process continues until no more lattice site in the system catches
fire. The energy dissipation (or outburst) is then completed. This is the
fastest process in the system and so the whole burning process is assumed to
take place in a single timestep.
\par
Finally, we define a ``burnable cluster'' as a collection of sites such that
all of them will catch fire in case any single one of them burns. This is a
useful concept when discussing the statistics of energy outburst.
\item {\em Energy Diffusion:} diffusion takes place in each timestep, i.e.
\begin{equation}
 E({\bf r})\rightarrow E({\bf r})+\Delta E({\bf r})\mbox{~~~~for~all~}{\bf r}
\end{equation}
where $\Delta E$ depends on the system configuration, the diffusion constant
$D$, the lattice size $\Delta L$, and the physical time $\Delta t$
corresponding to a cellular automaton timestep. But without lost of
generality, we can always rescale our time and space to make both $\Delta t$
and $\Delta L$ equal 1. Furthermore, the method we use to evaluate $\Delta
E({\bf r})$ is reported in the Appendix.
\end{enumerate}
\par \medskip
Various initial conditions, such as starting with all $E({\bf r}) = 0$, have
been used. However, the long term statistics is independent of the initial
conditions because both burning and diffusion erase the history of the system.
Although the model we have introduced is based on a two-dimensional square
lattice, generalizations to other dimensions and griding methods are straight
forward. More complete study of the model in various dimensions will be
reported in future works \cite{Continue}. Finally, a snapshot of a typical
system configuration immediately before a fire is triggered is shown in
Fig.~\ref{F:Snapshot}.
\section{Expectations And Simulation Results}
Before presenting the results of our numerical simulation, let us try to argue
qualitatively what we expect to see.
\subsection{Expectations}
There are three tuning parameters in our model: namely $E_{c1}$, $E_{c2}$, and
$D$. Large average dissipation size is a direct consequence of having large
burnable clusters immediately before the system catches fire. And large
burnable clusters can be obtained in either one of the following ways:
\begin{enumerate}
\item {\em Large Diffusion Constant:} diffusion spreads energy out and wipes
out information on where the energy packet is first introduced to the system.
Having a large $D$ means that immediately before a sudden energy dissipation,
the energy content in each site is approximately equal except for those few
sites where energy packets are recently introduced. So, we expect the whole
system to be covered by a single burnable cluster immediately before each
burning.
\item {\em Small Propagation To Triggering Thresholds Ratio:} in this case,
most of the sites have enough energy content to continue the propagation once
the burning begins. Once again, we expect to find a single burnable cluster
covering the entire system immediately before each burning \cite{Chau94}.
\end{enumerate}
\par \medskip
Using the same argument, small average dissipation size can be obtained when
diffusion is unimportant and when $E_{c2}$ is comparable to $E_{c1}$. For
every fixed $E_{c1}$ and $E_{c2}$, the average burnable cluster size
immediately before a burning increases with $D$. Therefore, starting from a
subcritical system (i.e. system with small burnable clusters only) with $D =
0$, a supercritical system (i.e. system with large burnable clusters only) can
be obtained simply by increasing the value of $D$. We also expect to find a
critical value of $D$, which is a function of $E_{c1}$ and $E_{c2}$, such that
burnable clusters of all sizes can be found immediately before burning. This
is the critical state in our model. Numerical experiments we have done
confirm our expectations, and we are going to report it in the coming
Subsection.
\subsection{The roles of diffusion}
In order to study the effects of diffusion, we vary the diffusion constant
$D$ from $10^{-6}$ to $10^{-2}$ in a $64\times 64$ lattice with $E_{c1}$ and
$E_{c2}$ equal 5.0 and 2.0 respectively. We found that the average size of
energy dissipation $\langle S\rangle$ increases as the diffusion constant $D$
increases (see Fig.~\ref{fig:rho}). Although we only report the system behavior
when it assumes the above parameters, the general behavior of the system using
different values of thresholds are unchanged.
\par \medskip
We observe that when $D \agt 3\times 10^{-3}$, both the energy dissipation
size $S$ and the time since last dissipation $T$ become very regular (see
Fig.~\ref{fig:prob_D}). This is the consequence of having a single burnable
cluster covering the whole surface just before the burning. Thus all the
energy available in the system is dissipated during the burning. As a result,
$S$ equals the total amount of energy dumped into the system since its last
dissipation which is in turn equal to $T$. The strong correlation between $S$
and $T$ for high diffusion system is shown in Fig.~\ref{fig:corr} (the circle
symbols). To study this correlation, we define the correlation coefficient
$\Gamma$ by:
\begin{equation}
 \Gamma = \frac{\langle ST \rangle - \langle S \rangle \langle T \rangle}{
 \sigma_S \sigma_T} \mbox{,}
\end{equation}
where $\sigma_S$ and $\sigma_T$ are the standard deviations of $S$ and $T$
respectively. It can be shown that $\Gamma$ varies from -1 to 1, and the value
of 1 (-1) implies that $S$ and $T$ are completely positive (negative)
correlated while 0 means that they are uncorrelated. The dashed line in
Fig.~\ref{fig:rho} shows that $\Gamma$ increases with increasing $D$
indicating that a strong positive correlation between $S$ and $T$ for system
with large diffusion constant. In this case, the distributions of energy
dissipation $P(S)$ and time interval $P(T)$ are the same, and both of them
show localized skewed peaks (see Fig.~\ref{fig:prob_D}).
\par \medskip
For a smaller $D$ ($5\times 10^{-4} \alt D \alt 3\times 10^{-3}$), the strong
correlation between $S$ and $T$ begins to break down. Weaker diffusion permits
greater fluctuation in the energy content at different sites. The system is
likely to catch fire before all the sites are connected. Hence not all energy
available in the system is dissipated, and some clusters may leave after a
burning. As shown in Fig.~\ref{fig:corr}, $(S,T)$ [the plus symbols] are
distributed around their mean values. In addition, if no burning is triggered
for sufficient long time, a system-wide dissipation can happen. $\Gamma$
decreases from 1 in this region as well. A consequence of the incomplete
burning is that the probability of having a triggering site in a small
burnable cluster is greatly increased. Events with small energy dissipation
size begin to appear and hence the correlation $\Gamma$ increases slightly as
$D$ decreases around $D = 6\times 10^{-4}$.
\par \medskip
As $D$ decreases further ($10^{-5} \alt D \alt 5\times 10^{-4}$), small
dissipation begins to dominate. In fact, $P(S)$ changes from localized peak
to power law and finally to exponential decay. In our simulation, we found
that the critical diffusion constant $D_{crit}$ to be $1.4\times 10^{-4}$.
Surely this critical value is a function of $E_{c1}$ and $E_{c2}$. For even
smaller $D$ ($D < 10^{-5}$), diffusion becomes insignificant. The heights of
nearby sites are nearly uncorrelated and hence $P(S)$ shows a very early
cutoff (i.e. an exponential decay). Fig.~\ref{fig:prob_D} depicts the
localized skewed peak, power law and exponential behaviors of $P(S)$ as a
result of different diffusion constant $D$.
\par \medskip
It should be noted that when $P(S)$ follows a power law, the corresponding
distribution of time interval between successive burning $P(T)$ decreases
exponentially for large $T$. It means that there is a characteristic time
interval for the occurrence of burnings even when scaling behavior is observed
in the distribution of dissipation sizes. It is because the burnable clusters
are spatially separated and the largest burnable cluster is only a portion of
the system. The total number of critical sites in the entire system and hence
the probability of triggering a burning are more or less the same all the
time. It is expected that this case has similar features with respect to the
one studied in the next section.
\par \medskip
\subsection{The roles of triggering and propagation thresholds}
In our simulations, we have not studied the effect of the thresholds on the
model in detail. However, we believe that the above behaviors would be also
observed when we vary $E_{c1}$ or $E_{c2}$ instead of $D$. In fact, when one
of the three parameters is fixed, we may observe any one of the above
behaviors by carefully choosing the other two parameters together with a
reasonable lattice size. For example, reducing $E_{c2}$ has similar effects as
increasing $D$ (see Fig.~\ref{F:Comparison}).
\section{Comparison with Forest-Fire Model}
Forest-fire model (FFM) was first introduced by Bak {\em et al.} as an example
of self-organized criticality (SOC) \cite{Bak90}. They considered a lattice of
cells where individual cell belongs to either one of the following states: a
green tree, a burning tree, and a void (i.e. no tree). In each timestep, a tree
grows on an empty cell with probability $p$, and a green tree burns if at least
one of its nearest neighbors contains a burning tree. Finally, a burning tree
becomes a void in the next timestep. Grassberger and Kantz \cite{Grass91}
later showed that the model is not critical in the limit $p \rightarrow 0$.
Later on, Drossel and Schwabl \cite{Dross92} modified the model by introducing
a lightning probability $f$ with which a green tree catches fire. This version
of FFM shows criticality in the sense that anomalous scaling laws are observed
provided that $(f/p)^{-\nu'} \ll p^{-1} \ll f^{-1}$ where $\nu' = 0.58$
for two-dimensional FFM \cite{Dross94}. In this
Section, we compare our model with the FFM introduced by Drossel and Schwabl.
\par \medskip
The parameters of our model should be chosen reasonably so that our model would
resemble the FFM. First of all, there is no diffusion in the FFM. So we require
$D \ll 1$. The propagation of fire in the FFM is always allowed and thus we
should have $E_{c2} \ll E_{c1}$. In 2-dimensional FFM, SOC behavior is claimed
provided that $(f/p)^{-0.58} \ll p^{-1} \ll f^{-1}$ \cite{Dross94}. It can be
interpreted as the time taken by a forest fire is much shorter than the time
taken by the growing of a tree, which is in turn much shorter than the time
interval of two successive lightning at the same site. In our diffusive and
dissipative cellular automaton model, the first criterion is satisfied
automatically because it takes only one timestep to burn down a cluster. For
the second criterion, we should set $E_{c1} \gg 1$ in order to reduce the
frequency of burning. On the other hand, $E_{c1}$ should not be too large, or
else the burnable clusters become too large and hence only large events are
observed. This is consistent with the FFM in which $f$ cannot be too small in
comparison to $p$.
After taking all the above criteria into consideration, we choose $E_{c1} = 6$,
$E_{c2} = 1$ and $D = 0$ with a $512\times 512$ periodic square lattice in our
simulation. Of course, other choices of parameters are possible (e.g. $E_{c1} =
6$, $E_{c2} = 0.1$ and $D = 10^{-7}$) and the results obtained are
qualitatively the same.
\par \medskip
In the FFM, a cluster of size $s$ is defined as a group of $s$ neighboring
sites with green trees. In contrast, the size $s$ of a burnable cluster in our
model is defined as the number of sites in that burnable cluster. In
Fig.~\ref{fig:NR}, we plot the distribution of cluster size $N(s)$ and the root
mean quadratic radius $R(s)$. We find that
\begin{equation}
 N(s) \sim s^{-\tau} \mbox{ with }\tau = 2.11 \pm 0.02 \mbox{,}
\end{equation}
and
\begin{equation}
 R(s) \sim s^{1/\mu} \mbox{ with }\mu = 1.87 \pm 0.02 \mbox{.}
\end{equation}
While the exponent $\tau$ is similar to the accurate results obtained by
Clar {\em et al.} \cite{Dross94} and Grassberger \cite{Grass93} for the FFM,
the exponent $\mu$ differs from the numerical value of $1.96 \pm 0.01$ obtained
by Clar {\em et al.} recently in their FFM simulations \cite{Dross94}.
\par \medskip
We then study the model with different $E_{c1}$ and some numerical results are
tabulated in Table~\ref{table:result}. The mean density $\rho$ of sites with
energy content greater than the propagation threshold $E_{c2}$ can be related
to the average burning size $\langle S \rangle$ by
\begin{equation}
 \langle S \rangle \sim |\rho_c - \rho |^{-\gamma}
\end{equation}
where $\rho_c$ is the mean density at critical point. Using the measured
results we have estimated that
\begin{eqnarray}
 \rho_c & = 0.410 \pm 0.002 \mbox{,} \\
 \gamma & = 2.57 \pm 0.01 \mbox{.}
\end{eqnarray}
Although the fitted value of $\gamma$ is different, $\rho_c$ is the same as
that of the FFM within numerical accuracy.
\par \medskip
The greatest difference between our model and the FFM is the probability
$p(s)$ of burning for a cluster of size $s$. In the FFM, $p(s)$ is proportional
to $s$ as all cluster sites are triggering sites in the sense that they are
ready to trigger a fire once lightning occurs. But in our model, only a few
percent of the sites in a burnable cluster are triggering sites.
Fig.~\ref{fig:prob} shows that the distribution of energy dissipation $P(S)$
scales as $S^{-\alpha}$ with an $\exp (-S/S_{\xi})$ cutoff. We also plot the
distribution of burning cluster size $D(s)$ and find that within the accuracy
of our simulation, $P(S)$ and $D(s)$ share the same exponent in the scaling
region. Thus we have $s \sim S$ and hence
\begin{equation}
 D(s) \propto s^{-\alpha}\exp(-s/s_{\xi}) \mbox{ with }\alpha = 0.85
 \pm 0.01 \mbox{,}
\end{equation}
while it is known that $\alpha \approx 1.0$ for the FFM. Further numerical
experiments show that $\alpha$ decreases as $D$ increases in our model
(see Fig.~\ref{F:Comparison}). In conclusion, although our model is similar
to that of the FFM, they belong to two different universality classes.
\par \medskip
Although $\alpha$ depends on the value of $D$, we find within numerical
errors that the value of $\alpha$ does not change with $E_{c1}$. In
Fig.~\ref{fig:h_c1} one can see that different $E_{c1}$ affects the range of
scaling region but not the scaling exponent of $P(S)$. It is easy to realize
that for a larger $E_{c1}$ it takes a longer time to trigger a burning and
allows the growth of larger clusters. In order to prove the fall off of $D(s)$
for large $s$ is not due to the finite size of the system, simulations using
larger lattice size have been done and the same spectra of $D(s)$ (in terms of
the number of sites $s$) are obtained. Furthermore, the distribution of time
interval between successive burnings $P(T)$ shows an exponential decay as
discussed in Section~II. We conclude in our simulation that no system-wide
dissipation occurs.
\par \medskip
In addition to the RMS radius, we define a maximum elongation $d(s)$ which is
the maximum distance between any two sites in a cluster of size $s$ to
investigate the shape of the clusters. We found that
\begin{equation}
 d(s) \sim s^{1/\mu'}\mbox{ with }\mu' = 1.82 \pm 0.02 \mbox{.}
\end{equation}
Fig.~\ref{fig:NR}b tells us that for small $s$, both $R(s)$ and $d(s)$ deviate
from a straight line. This is because of the discreteness of the lattice. For
larger value of $s$, however, they follow power laws of similar exponents and
the ratio $d(s)/R(s)$ increases to a constant value around 2.8 which is very
close to the value of a 2-dimensional compact
circular object (which equals $\sqrt{8}$). This concludes that small burnable
clusters are generally irregular and asymmetric objects due to the random
fluctuation of their formation. As the size of a burnable cluster increases,
clusters become more and more regular and spherical. The ratio $d(s)/R(s)$
shows that in the average the clusters are more or less compact objects for
large $s$. On the other hand, the values of $\mu$ and $\mu'$ suggest that the
clusters may have fractal structures with a dimension around $1.9$. However, at
this stage we cannot determine the fractal dimension with accuracy because
there is no a general efficient method to estimate it and also our simulation
is not close enough to the critical point of the system. It is interesting to
note that when we choose a random site near the center of mass of a large
cluster, the average local density of the cluster around that site is close to
the site percolation threshold $0.59$. We believe that there is a close
relationship between site percolation and our model (or the FFM) and further
investigation is needed to draw any conclusion.
\section{Conclusions And Outlook}
In this paper, we introduce a cellular automaton model for diffusive and
dissipative systems, and we have described its behaviors in various different
locations in the parameter space. First, we study the effect of diffusion and
find that $P(S)$ may follow a localized skewed peak, power law, or exponential
decay. Then we consider the case without diffusion and compare it with the
famous forest-fire model.
\par \medskip
With a careful choice of parameters and geometry, this model is able to
describe qualitatively the behavior of a number of physical systems, including
the behaviors of type-I X-ray bursters (details can be found in Reference
\cite{Chau94}) and CO$_2$ gas outbursts in some crater lakes (details can be
found in Reference \cite{Lakes}). The following is a brief description of the
above two physical systems and the reason why our cellular automaton model can
be used to describe their behaviors.
\par \medskip
Type-I X-ray bursts are results of thermonuclear explosion on the surface of an
accreting neutron star. Nuclear fuel, usually making up of hydrogen and helium,
is deposited on the neutron star surface. Whenever the local density of nuclear
fuel is greater than a threshold value, which is determined by the specific
nuclear reaction involved, a thermonuclear runaway takes place. The nuclear
reaction can spread to its neighbors if they contain sufficient amount of
nuclear fuel. It results in a transient flash in the X-ray band, called a
type-I X-ray burst. Type-I bursts are different from the more frequently
repeating type-II bursts, which are believed to be results of accretion disk
instability \cite{Lewin}. It is easy to estimate also that the material
diffusion timescale in this problem is so long as compared to the typical time
interval between two successive bursts. Thus material diffusion is not
important in this problem. After taken into account for the spatial fluctuation
in the accretion process, the system can be described using our cellular
automaton model on a spherical surface with $D = 0$. In addition, the values of
$E_{c1}$ and $E_{c2}$ can be determined in principle once the specific nuclear
reaction, cooling process, and the typical mass of an accreting blob are given.
And the burst statistics predicted by our model is consistent with
observations \cite{Chau94}.
\par \medskip
Another application of the diffusive and dissipative model is the gas outburst
statistics in some crater lakes. CO$_2$ gas is injected to the bottom of the
lake by some natural processes \cite{Lake}. Note that the solubility of CO$_2$
in water increases with pressure (and hence the depth of water). So once the
CO$_2$ concentration in the bottom of the lake becomes supersaturated, gas
bubble will form and it will drive the water around it to move up causing
further degassing. A catastrophic gas outburst is resulted. Taking into
account the horizontal water flow in the lake, we can describe the outburst of
a crater lake using our model with a large $D$, and hence rather regular
gas outburst is expected.
\par \medskip
Further investigation of this model in other spatial dimensions, behavior of
the model at criticality as a function of diffusion constant $D$, and other
possible applications of the model will be reported in future work
\cite{Continue}.
\newpage
\section*{Appendix: Method Of Handling Diffusion}
The energy diffusion equation in continuous flat two-dimensional space is
\begin{equation}
\frac{\partial \rho}{\partial t} =  D \left( \frac{\partial^2 \rho}{\partial
 x^2} + \frac{\partial^2 \rho}{\partial y^2} \right)
\end{equation}
where $\rho$ is the energy density.
In our cellular automaton model, we approximate the above equation using
the finite difference:
\begin{equation}
 E({\bf r},t+\Delta t) = D \frac{\Delta t}{\Delta L^2} \sum_{\bf r'} \left[
 E({\bf r'},t) - E({\bf r},t) \right] + E({\bf r},t) \mbox{~~~~~~for~all~}
 {\bf r} \label{eq:fd-d}
\end{equation}
where $E({\bf r},t)$ is the energy content of site $\bf r$ are time $t$, and
the sum is over all the nearest neighbors of ${\bf r}$. This is possible
since the areas of lattice sites are the same. The value of $D$ is coupled
with the system size and the timestep employed. Eq.~(\ref{eq:fd-d})
is a good approximation to the actual diffusion process in continuous
spacetime if $D \Delta t / \Delta L^2 \ll 1$. This criteria is satisfied in
all the cases we have reported in this paper. In the event that $D$ is too
large, an adaptive integration scheme to calculate $E(r,t+\Delta t)$ is
employed.
\par \medskip
The diffusion constant is an invariant under any rescaling of the system. This
is true provided that the changes in $\Delta L$ and $\Delta t$ are correlated.
That is, $\Delta t \longrightarrow \lambda^2 \Delta t$ whenever $\Delta L
\longrightarrow \lambda \Delta L$.
\acknowledgements{We would like to thank B. Drossel and the anonymous referee
for their useful comments. This work is supported by UPGC grant of Hong Hong,
DOE grant DE-FG02-90ER40542 and NSF grant AST-9315133 of the U.S.A.}

\begin{table}
 \begin{tabular}{cccc}
  $E_{c1}$ & $\Gamma$ & $\langle E \rangle$ & $\rho$ \\ \hline
  4 & $-0.011\pm 0.006$ & $53.0\pm 0.4$ & $0.3343\pm 0.0005$ \\
  5 & $-0.008\pm 0.006$ & $178\pm 2$ & $0.363\pm 0.001$ \\
  6 & $+0.012\pm 0.006$ & $643\pm 6$ & $0.381\pm 0.001$ \\
  7 & $+0.02\pm 0.01$ & $2440\pm 40$ & $0.393\pm 0.001$ \\
 \end{tabular}
 \hspace{0.25in}
 \caption{Some numerical results on a $512\times 512$ lattice with
  $E_{c2}=1$ and $D=0$.}
 \label{table:result}
\end{table}
\begin{figure}
 \caption{A greyscale snapshot of a $512\times 512$ lattice with $E_{c1}=6$,
  $E_{c2}=1$, and $D=0$. The higher the value of $E$, the darker the dot is.}
 \label{F:Snapshot}
\end{figure}
\begin{figure}
 \caption{The variation of the mean dissipation size $\langle S \rangle$
  (solid line) and the correlation coefficient $\Gamma$ (dash line) against
  the diffusion constant $D$. All the simulations are done in a $64\times 64$
  lattice with $E_{c1} = 5.0$ and $E_{c2} = 2.0$ respectively. The same set of
  parameters is used in Figs.~\protect\ref{fig:prob_D} and
  \protect\ref{fig:corr}.}
 \label{fig:rho}
\end{figure}
\begin{figure}
 \caption{Distributions of (a) energy dissipation size $S$, and (b) time
  interval between successive energy dissipation $T$ for different values of
  $D$.}
 \label{fig:prob_D}
\end{figure}
\begin{figure}
 \caption{Correlation of the time since last energy dissipation $T$
  and the size of energy dissipation $S$ for different values of $D$.}
\label{fig:corr}
\end{figure}
\begin{figure}
 \caption{The distribution of energy dissipation size of a system with
  $D=5\times 10^{-5}$ and $E_{c2}=2$ (solid line).  The effect of increasing
  $D$ to $2\times 10^{-4}$ (dotted line) is qualitatively similar to that of
  reducing $E_{c2}$ to $1$ (dashed line).}
 \label{F:Comparison}
\end{figure}
\begin{figure}
 \caption{The distribution of (a) cluster size $N(s)$; (b) RMS radius
  $R(s)$ and maximum elongation $d(s)$.}
 \label{fig:NR}
\end{figure}
\begin{figure}
 \caption{A plot of (a) $P(S)$ against $S$; and (b) $D(s)$ against $s$ with
  $E_{c1}=6$, $E_{c2}=1$ and $D=0$ in a $512\times 512$ lattice.  A scaling
  region from $10$ to $10^{3}$ is observed with an exponent of $-0.85$.}
 \label{fig:prob}
\end{figure}
\begin{figure}
 \caption{Distribution of energy dissipation size $P(S)$ with $E_{c1}=$7
  (solid line), 6(long dash line), 5(dotted line) and 4 (med dash line)
  for a $512 \times 512$ lattice.  $E_{c2} = 1$, and $D = 0$. As since
  in the $S$ vs. $P_{E_{c1}} (S) / P_{E_{c1}} (10)$ plot, the scaling
  region increases with increasing $E_{c1}$. However, the scaling exponent
  is independent of $E_{c1}$.}
 \label{fig:h_c1}
\end{figure}
\end{document}